\documentclass[%
 reprint,
 amsmath,amssymb,
 aps,
]{revtex4-1}

\usepackage{graphicx}
\usepackage{epstopdf}

\usepackage{dcolumn}
\usepackage{bm}
\usepackage{xcolor}

\begin{document}

\preprint{APS/123-QED}

\title{Ultrafast strain-induced charge transport in semiconductor superlattices}
 
 \author{F.~Wang$^{1,2}$, C.L.~Poyser$^2$, M.T.~Greenaway$^1$, A.V.~Akimov$^2$, R.P.~Campion$^2$, A.J.~Kent$^2$, T.M.~Fromhold$^2$, and A.G.~Balanov$^1$}
 \affiliation{$^1$Department of Physics, Loughborough University, Loughborough LE11 3TU, United Kingdom\\
$^2$School of Physics and Astronomy, University of Nottingham, Nottingham NG7 2RD, United Kingdom\\
}

\date{\today}%
\begin{abstract}

We investigate the effect of hypersonic ($>$ 1 GHz) acoustic phonon wavepackets on electron transport in a  semiconductor superlattice. Our  quantum mechanical simulations demonstrate that a GHz train of picosecond deformation strain pulses propagating through a superlattice can generate current oscillations whose frequency is several times higher than that of the strain pulse train. The shape and polarity of the calculated current pulses agree well with experimentally measured electric signals. The calculations also explain and accurately reproduce the measured variation of the induced current pulse magnitude with the strain pulse amplitude and applied bias voltage. Our results open a route to developing acoustically-driven semiconductor superlattices as sources of millimetre and sub-millimetre electromagnetic waves.

\end{abstract}

\maketitle

\section{Introduction}

The development of  picosecond techniques for generating ultra-fast coherent phonons \cite{Thomsen1985, Eesley1987} has provided researchers with new efficient acoustoelectric tools for controling charge transport in semiconductor nanostructures (see \cite{Akimov2017} for review). For example, acoustoelectric effects have been proposed for the generation \cite{Fowler2008,Greenaway2009,Greenaway2010, Apostolakis2017}, heterodyne-mixing \cite{Heywood2016}, and detection \cite{Batista2007,Kent2006} of high-frequency electromagnetic waves. Acousto-electronic pumps have also been used to control quantum phenomena in nanostructures and quantum dots \cite{Talyanskii1997,Cunningham2000,Ahn2007,Gell2009,Young2011,Poyser2015a}, with recent experiments demonstrating that ultrafast acoustic pulses can dramatically enhance the emission output of a quantum-dot cavity  \cite{Brue2012,Wigger2017}.   Also, semiconductor nanostructures can serve as active media for transforming THz acoustic waves into sub-mm electromagnetic radiation \cite{Stanton2003, Armstrong2009}. It has also been recently shown that GHz-strain pulses can be rectified by a semiconductor superlattice (SL) producing a DC response similar to that due to rectification of microwaves \cite{Poyser2015}.  
Despite this progress in understanding the GHz-THz acoustoelectric phenomena, the induced high-frequency charge transport and the corresponding electronic response to changing the parameters of the strain pulses have not yet been widely considered. In particular, it is still unclear how the coherent acoustic/strain stimuli affect the quantum tunneling of the charge and the resulting current flow dynamics.

 In this paper, we present a quantum-mechanical theory of electron transport in a biased weakly-coupled SL driven by a coherent phonon wavepacket in the form of a train of ultrafast (picosecond) strain pulses. We solve numerically the Schr\"{o}dinger equation with a time-dependent potential generated by propagating deformation pulses. Our simulations show that the polarity and the magnitude of the generated current response depend on the amplitude of the strain pulses and on the applied bias voltage. The results of our numerical modelling agree well with the experimental measurements of the electrical signals from the SL that we present. We also discuss how the high-frequency electronic response of the SL to coherent acoustic phonons differs fundamentally from the effects of electromagnetic radiation on the SL. This difference is due to the finite propagation time of the phonons through the SL.  Finally, our theoretical analysis predicts that the acoustic stimuli can induce current oscillations in the device, with frequencies several times higher than that of the strain pulse train itself. The high-frequency oscillations of the current reflect those of the drifting electrons within the alternating acoustic potential propagating along the SL.

The paper has the following structure. In Section \ref{sec:mod} we present our quantum model and numerical simulations of electron transport in a voltage-biased SL driven by strain pulses. 

We present the results and related discussion in Section \ref{sec:res}. Section \ref{sec:con} provides conclusions and outlook.
 
\section{Theoretical model}
\label{sec:mod}
\begin{figure*}[t]
\centering
\includegraphics[width=0.5\textwidth]{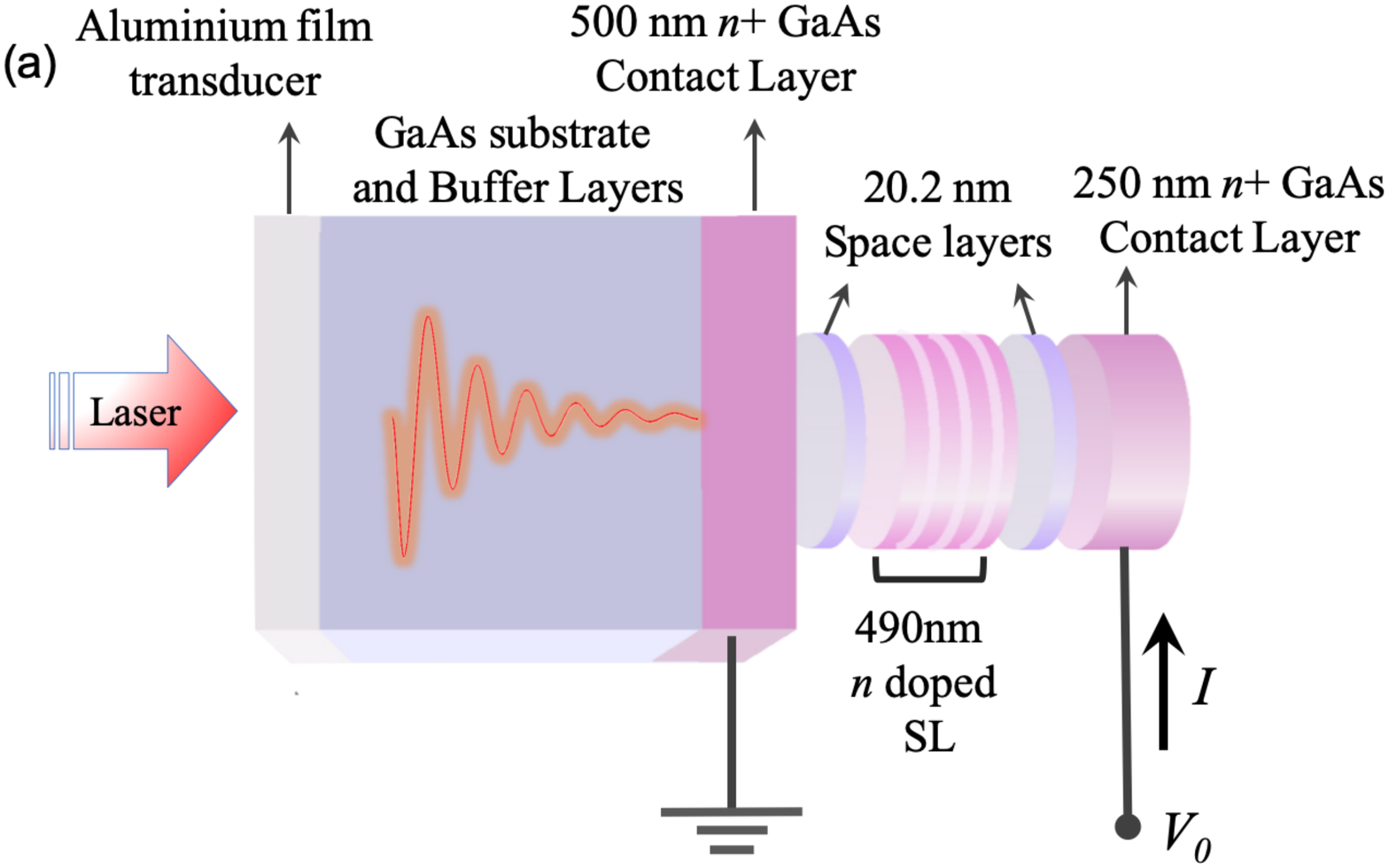}
\includegraphics[width=1.0\columnwidth]{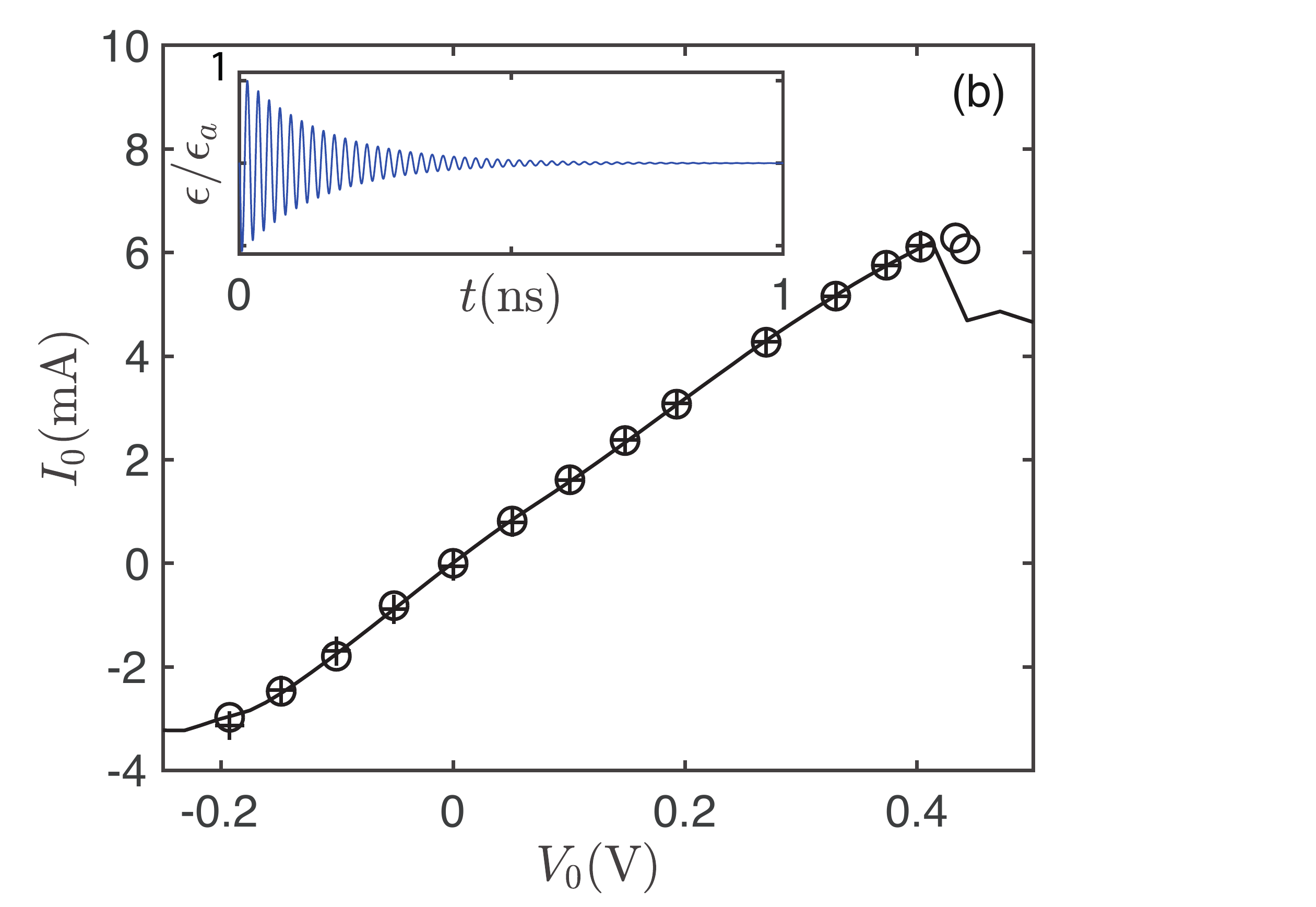}
\caption{\label{fig:static} (a) Schematic diagram of our device, which includes (towards the right) the weakly coupled SL, contacts and spacer layers. The optoelastic Al film transducer (left) is deposited on the side of the GaAs substrate opposite to the device. (b) The current-voltage characteristics, $I_0(V_0)$, of the device calculated numerically  (open circles), approximated by a cubic polynomial (crosses) and measured experimentally  (solid curve); inset: the numerically simulated time-realisation of strain pulses with a frequency of 46 GHz generated by the Al film transducer, $\epsilon_a$ is the maximum strain within the train of pulses. }
\label{fig:schema}
\end{figure*}

In our study, we model the experiment, described in detail in Ref. \cite{Poyser2015}.  
The device is schematically pictured in Fig.~\ref{fig:schema} (a). The SL region of length $L=490$ nm is formed by 50 GaAs and AlAs layers with thicknesses of 5.9 and 3.9 nm respectively, giving a lattice period of $d = 9.8$ nm. The bias voltage $V_0$ is applied to the ohmic contacts. The train of strain pulses is generated by femtosecond laser pulses incident on an Al film with thickness 30 nm deposited on the reverse to the SL  GaAs substrate. We estimate the strain amplitude value using previous experimental measurements \cite{Akimov2007}, where it has been shown that the maximum strain amplitude $\epsilon_a=10^{-4}$ corresponds to a laser pulse energy density of $\approx1$ mJ cm$^{-2}$.
The time-dependent current $I(t)$
was measured  using a digitizing oscilloscope with a 12.5 GHz bandwidth. All measurements were performed at a lattice temperature of $T=5$ K.

To calculate charge transport in the SL we solve numerically the time-dependent Schr\"odinger equation 
\begin{equation}
\label{eq:sch}
-i\hbar\frac{\partial \psi(x,t)}{\partial t} = \hat{H}\psi(x,t), 
\end{equation}
for the electron wavefunction $\psi(x,t)$, taking the Hamiltonian to be

\begin{equation}
\hat{H} = - \frac{{{\hbar ^2}}}{{2{m}}}\frac{\partial ^2}{\partial x^2} +U_{SL}(x) - eV_0x + {U_{a}(x,t)}, \nonumber
\end{equation}
where $m=0.067 m_e$  ($m_e$ is the electron mass) is the electronic effective mass in GaAs, $e$ is the electron charge, $V_0$ is the bias voltage applied to the SL, $U_{SL}(x)$ is the SL potential, and $U_{a}(x,t)$ is 
the time-dependent potential energy field generated by the deformation pulses [see inset in Fig.~\ref{fig:schema} (b)],
which includes reflections of the acoustic wave from the boundaries and is given, together with the corresponding parameters, by Eq.~(7) in \cite{Poyser2015}. This potential propagates along the SL with the speed of sound  in GaAs/AlAs layers $v_s\approx5000$ m/s. To integrate Eq. (\ref{eq:sch}) we use the Crank-Nicolson method, adapted for General Purpose computing on Graphics Processing Units (GPGPU) using the CUDA platform.  From $\psi(x,t)$, we calculate the expectation value of the electron displacement $\left< x \right>$.
With no acoustic wave ($U_a=0$), a non-zero bias $V_0 \ne 0$ induces Bloch oscillations with a frequency proportional to $V_0$ and an amplitude inversely proportional to $V_0$ \cite{Bloch1929}.

To calculate the DC current, $I_0$, in the SL with no acoustic stimulus applied, i.e. $U_a=0$,
we introduce the drift velocity $v_d$ of the electrons using the Esaki-Tsu formalism \cite{Esaki1970}

\begin{equation}
\label{eq:i-dc}
v_d=\delta \int_{0}^{\infty} \frac{dt}{\tau} \left<v(t)\right> e^{-t/\tau},
\end{equation}
which takes into account the effect of electron scattering. Here, $\tau=\tau_i\delta=0.47$ ps and $\delta=\sqrt{\tau_e/(\tau_i+\tau_e)}=12.9$ are parameters that depend on the inelastic, $\tau_i$ and elastic, $\tau_e$, scattering times \cite{Fromhold2004}. The electron velocity $\left<v(t)\right>$ is found by averaging $d\left<x\right>/dt$ over different initial phases of the Bloch oscillations in the interval $[0,2\pi]$. The DC electric current in the SL is $I_0=\pi r^2 n_{e} e v_d$, where $r=50 \mu$m  is the radius of the SL mesa and $n_e=10^{17}$ cm$^{-3
}$ is the doping density. The effect of the ohmic contacts is taken into account by setting $V_0=V_{SL}-I_0R_c$, where $V_{SL}$ is the voltage drop across the SL and $R_c=54.7$ $\Omega$ is the contact resistance. 

The time-dependent current $I(t)=\pi r^2 n_{e} e v_d(t)$  generated
by the voltage-biased SL driven by the strain pulses is calculated using the time-dependent drift velocity \cite{RENK1998}
\begin{equation}
\label{eq:i-ac}
v_d(t)= \int_{-\infty}^{t} \frac{dt_0}{\tau} \left<v(t)\right> e^{-(t-t_0)/\tau}.
\end{equation}
To study the response of the SL to strain stimuli, we analyse the incremental current $\delta I(t)=I(t)-I_0$.


\begin{figure*}[pt]
\includegraphics[width=\textwidth]{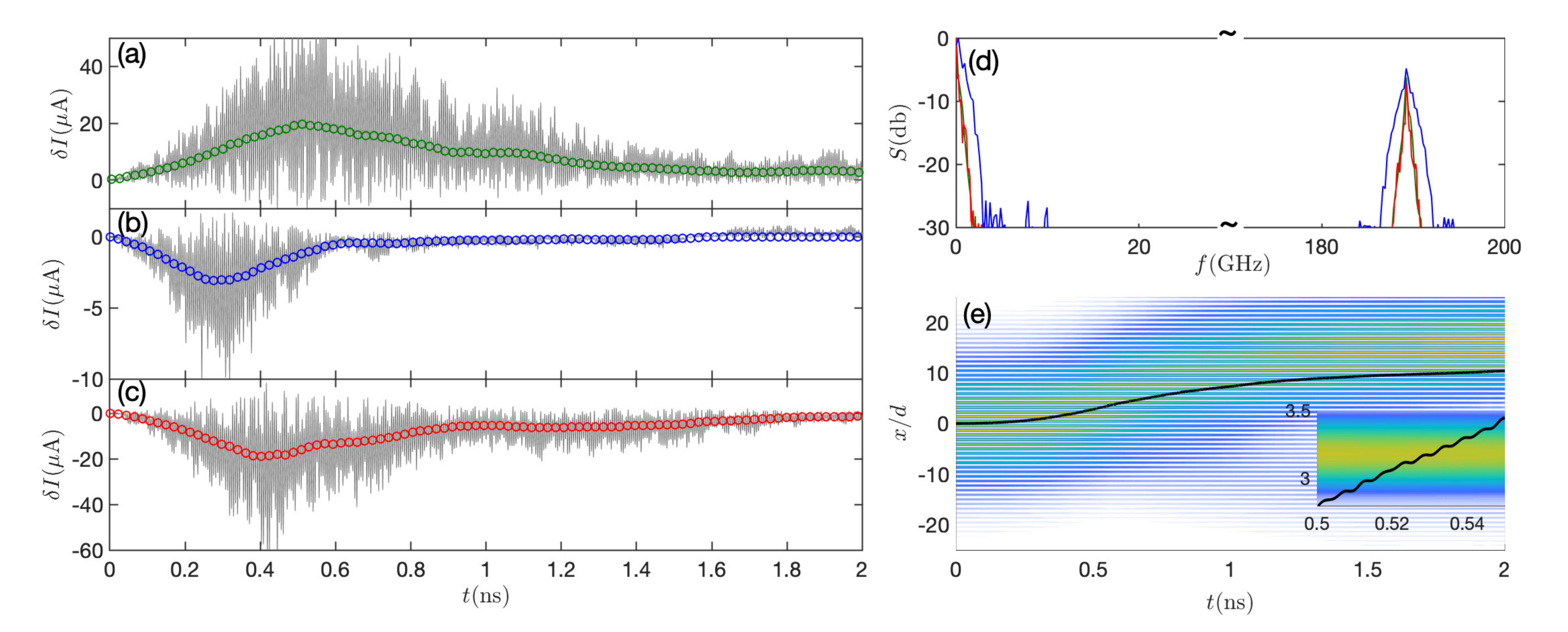}

\caption{(a-c) Time-dependent incremental current $\delta I(t)$ calculated for $V_0=$ (a) -100 mV, (b)  0 mV, and (c) 100 mV, taking $\epsilon_a=1.5\times10^{-4}$. Gray curves show the signal and colored curves are calculated using the moving average method with a 12.5 GHz cut off frequency. (d) Fourier spectra $S$ of the (gray) signals in (a-c). The colors of the Fourier spectra correspond to those of the respective averaged signals in (a),(b) and (c). (e) Color maps: evolution of the probability density (yellow = high) of the acoustically driven electron wavepacket calculated for $V_0=-100$ mV, as in (a), shown enlarged in the lower right inset. Black curves: averaged electron trajectory $\langle x\rangle(t)$, revealing the drift and superimposed oscillations (inset) also seen in the wavepacket dynamics.}
\label{fig:iac}
\end{figure*}

\section{Results and discussion}
\label{sec:res}

The solid black curve in  Fig. \ref{fig:static}(b) shows the current-voltage characteristic, $I_0(V_0)$, measured for the SL used in our experiments, and whose parameters (see  Fig. \ref{fig:static}(a) and Section \ref{sec:mod}) were used in the model (1)-(3) and resulting numerical simulations.
 For $V_0$ between -0.2 and +0.4 V, $I_0$ grows monotonically with increasing $V_0$. However, when $V_0>0.4$ V an electric instability results from the onset of negative differential conductivity (NDC) \cite{Wac02}, which induces propagating charge domains and associated current self-oscillations \cite{Kastrup1996,Greenaway2009a}. In order to avoid these propagating charge domains affecting and complicating our measurements and analysis, we ensure that $|V_0| \leq 300$ mV. The measured $I_0(V_0)$ characteristic was used to determine $\tau$, and $\delta$ (see Section \ref{sec:mod}) by fitting the corresponding calculated curve to it. As shown in Fig. \ref{fig:static}(b), the model (\ref{eq:sch}) -- (\ref{eq:i-dc}) (empty circles) reproduces the measured $I_0(V_0)$ curve accurately.

Typical $\delta I(t)$ current pulses calculated numerically for $\epsilon_a=1.5\times10^{-4}$ and  $V_0 = $ 100~mV, 0~mV and -100 mV are shown by the gray solid curves in Figs. \ref{fig:iac} (a), (b) and (c), respectively.  Our calculations reveal fast oscillations with a frequency that is several ($\sim 4$) times larger than the 46 GHz frequency of the acoustic oscillations.  Notably, our results are qualitatively different from the case of a SL driven by an electromagnetic wave \cite{Wacker1997b,Wac02}, which responds only to the fundamental frequency (and harmonics) of the AC driving field. Figure \ref{fig:iac} (d) shows the Fourier spectra calculated for the signals in Fig. \ref{fig:iac} (a-c). All three spectra reveal a sharp peak at a frequency ($190$ GHz) that is far higher than the frequency of the picosecond pulse train ($46$ GHz). 
 Our quantum simulations show that this originates from slow propagation of the strain pulses, which, in contrast to the case of electromagnetic driving (see Appendix), cannot be neglected. Figure \ref{fig:iac} (e) illustrates the dynamics of the wavepacket $|\psi^2(x,t)|$ as it travels through the SL. Initially, it is accelerated by the acoustic wave incident on the SL, but after $\sim 1.5$ ns the wavepacket slows due to the decay of the acoustic pulse (see inset in Fig. \ref{fig:schema} (b)). However, for $t\leq 1.5$ ns, the wavepacket travels with almost constant mean speed ($\approx$ 17.8  $\mathrm{ m/s}$), whilst also demonstrating small superimposed high-frequency oscillations, see the averaged electron trajectory $\langle x\rangle(t)$  in Fig. \ref{fig:iac} (e) and its inset. 

Similar behavior was previously predicted in \cite{Greenaway2010}, where it was also shown that electron miniband transport induced by a plane acoustic wave can up-convert the frequency of the wave, into faster electron oscillations, by the factor     

\begin{equation}
\label{eq:beta}
\beta=(\epsilon_a D \Delta_{SL} / \pi )^{1/2}(d/\hbar v_s),
\end{equation}
where $\Delta_{SL}$ is the width of the SL miniband and $D\approx10$ eV is the deformation potential for GaAs \cite{Adachi1992, Gor1993}.
Within a semiclassical picture, these fast oscillations can be associated with pendulum-like motion of electrons in the propagating acoustic (strain) potential \cite{Apostolakis2017}.
For our experimental SL, we estimate $\Delta_{SL}\approx3.4$ meV, which yields $\beta=3.79$. This value agrees well with the 4.1 up-conversion factor demonstrated in Fig. \ref{fig:iac} (d).

\begin{figure}[t]
\includegraphics[width=\columnwidth]{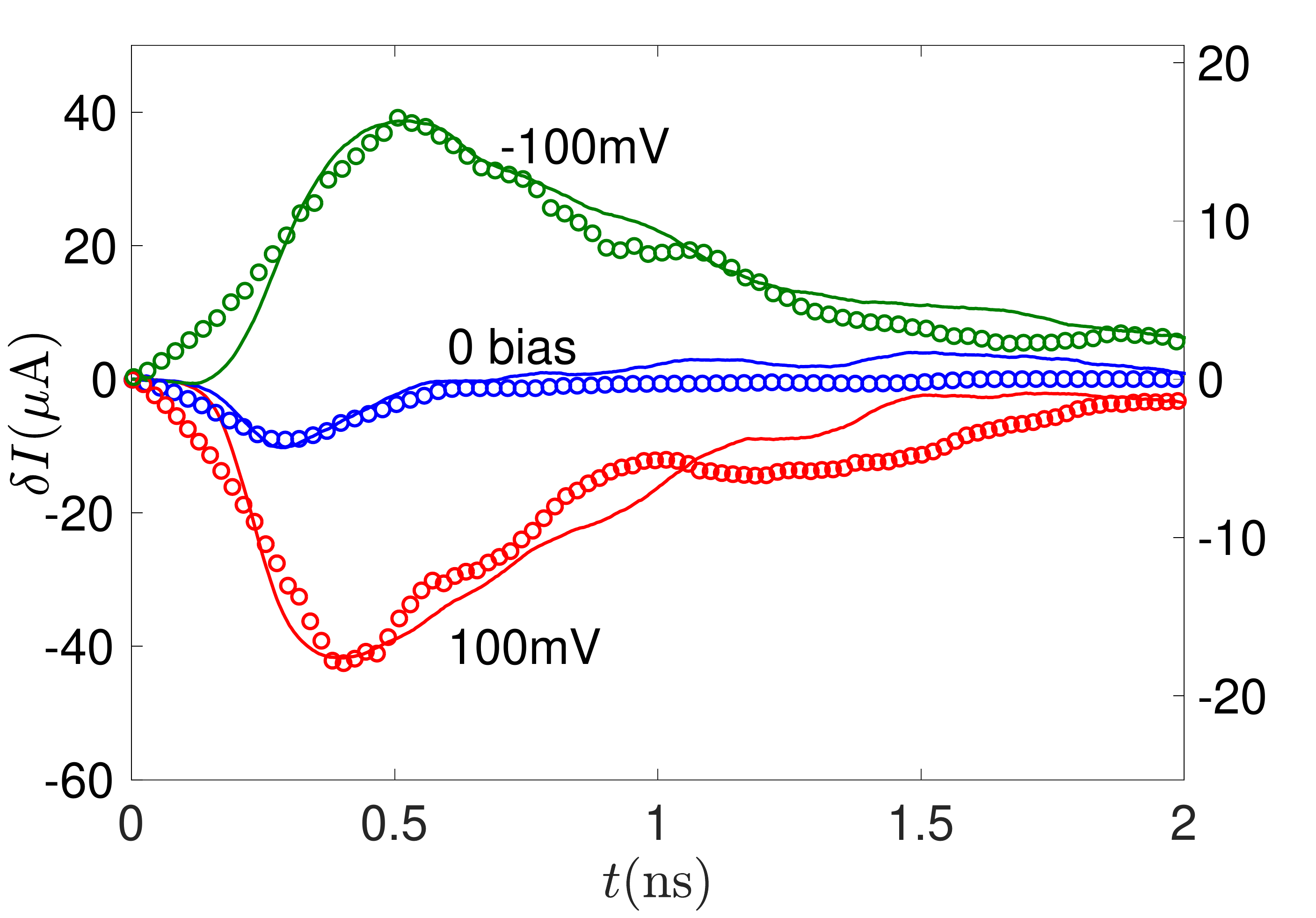}

\caption{Typical current pulses $\delta I(t)$ calculated numerically (open circles, right axis) and experimentally measured (solid curves, left axis) for $V_0=+100$ mV (red), -100 mV (green) and 0 (blue), $\epsilon_a=1.5\times10^{-4}$. }

\label{fig:iac-exp}
\end{figure}

        This theoretical analysis is supported by our experiments. Comparison of the time-dependent incremental current $\delta I(t)$ calculated numerically using Eqs.(\ref{eq:sch}) -- (\ref{eq:i-ac}) and measured experimentally for $\epsilon_a=1.5\times10^{-4}$ is presented in Fig. {\ref{fig:iac-exp}}. The green, blue and red empty circles (simulations) and solid curves (measurements) correspond to 
      $V_0=-100$ mV, 0 mV, and 100 mV, respectively. Since the measurements of $\delta I(t)$ were made using an oscilloscope with a 12.5 GHz bandwidth, we apply to our calculated $\delta I(t)$ signals a moving average filter with a cut-off frequency equal to the bandwidth of the oscilloscope, see also empty circles in Fig. \ref {fig:iac} (a-c). The calculated $\delta I(t)$ (empty circles, right axis) characteristics exhibit clear qualitative agreement with both the shape and polarity of the experimentally measured $\delta I(t)$ curves (solid line, left axis).
      Interestingly, the polarity of the response current is opposite to the polarity of the bias. The start of the leading edge of $\delta I(t)$ corresponds to the arrival of the strain pulse at the contact.  The subsequent decay of $\delta I(t)$ is determined by the duration of the strain pulses \cite{Poyser2015}, see inset of Fig. \ref{fig:static}(b), and their reflections from the boundaries of the SL. The blue curves show that a non-zero $\delta I(t)$ response is generated even when $V_0=0$, demonstrating the direct effect of the strain pulses on charge transport in the SL, which is amplified further when $V_0\pm100$ mV.  The scaling factor between the simulated (right axis) and measured (left axis) current pulses may originate from simplifications made in our model, in particular, omission of complex high-frequency charge dynamics in the contact regions \cite{Balanov2012,Maksimenko2015} and parasitic frequency-dependent reactance of the contacts and leads \cite{Alexeeva2012}.

\begin{figure}[t]
\includegraphics[width=1.\columnwidth]{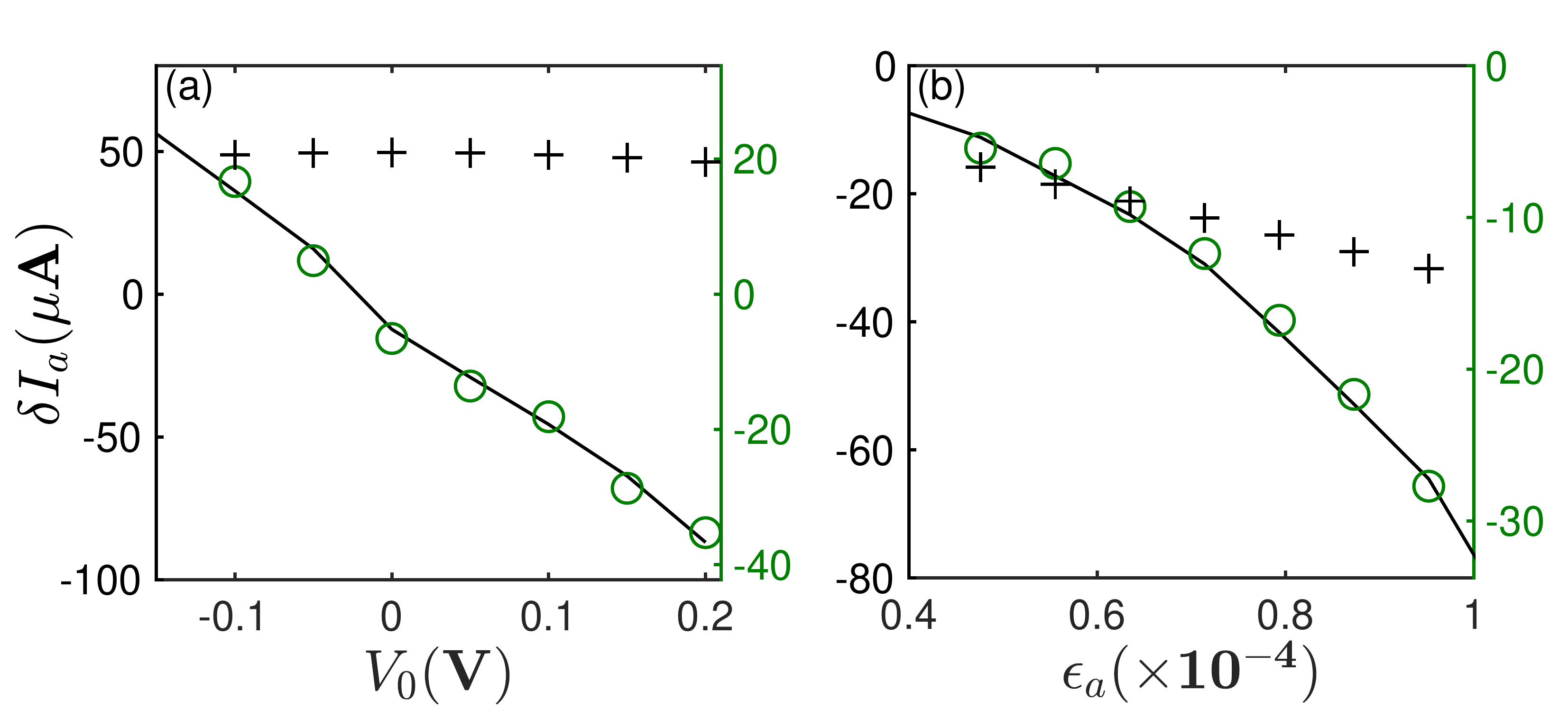}
\caption{Calculated (symbols, right axis) and measured (solid lines, left axis) variation of current pulse amplitude $\delta I_a$ with applied bias $V_0$ (a) and strain pulse amplitude $\epsilon_a$ in the train (b). Open circles are numerical calculations. Crosses are the values [divided by factor 20 in (a), and -20 in (b)] calculated from an analytical model developed to describe the response of the SL to irradiation by an electromagnetic wave (see Appendix and Eq.6).}
\label{fig:di-dep}
\end{figure}

Next, we study how the peak deformation stain $\epsilon_a$ and bias voltage $V_0$, applied to the SL, affect the amplitude $\delta I_a$ of the generated current pulses $\delta I(t)$, where $\delta I_a$ is defined as the maximum/minimum of $\delta I(t)$ in the case of positive/negative polarity of the pulse (see Fig. {\ref{fig:iac-exp}}).
Figure \ref{fig:di-dep} summarises our analysis of the current response to variation of $V_0$ and $\epsilon_a$, where the calculated $\delta I_a$ values are shown after filtering with a bandwidth of 12.5 GHz. The empty circles in Figs. \ref{fig:di-dep}(a) show $\delta I_a$ values calculated numerically for $\epsilon_a=1.27\times10^{-4}$ and various $V_0$ values. The positive or negative values of $\delta I_a$ reflect the polarity of the current pulses, see also Fig. \ref{fig:iac-exp}. 
Increasing the bias magnitude $|V_0|$ leads to proportional change in  $\delta I_a$. The polarity of the response current  changes at $V_0\approx$ 50 mV, reflecting the fact that even in the absence of an applied bias, i.e. $V_0$=0, the acoustic excitation generates a current response with negative polarity (see Fig. {\ref{fig:iac-exp}}).
The open circles in Fig. \ref{fig:di-dep}(b) show the numerically calculated $\delta I_a(\epsilon_a)$ dependence for $V_0=300$ mV. These calculations show that the magnitude of the current pulses grows nonlinearly and monotonically as the strain stimuli become stronger, and preserves the polarity of the current pulse defined by the bias $V_0$. 

      The experimentally measured $\delta I_a(V_0)$ and $\delta I_a(\epsilon_a)$ variations are presented in Fig. \ref{fig:di-dep}(a) and (b) by solid lines. In all cases, the numerically calculated values demonstrate excellent agreement with the measured characteristics, thus confirming the the validity of our numerical model.
     
     Finally, we compare the response of the SL to the acoustic excitations with known results on the rectification and detection of electromagnetic waves by semiconductor tuneling devices \cite{Tucker1979,Wacker1997b,Wac02}, based on a theoretical framework that we discuss briefly in the Appendix. To facilitate this comparison, we calculated the amplitude of the time-dependent response of the SL to electromagnetic irradiation (photons), which produces a potential with the same frequency and amplitude as the deformation pulses (coherent phonons) in our study (see Eq. (\ref{eq:iac_p}) in the Appendix).  We note that since the electromagnetic waves propagate through the lattice at the speed of light, and their wavelength in GHz range significantly exceeds the SL length $L$, propagation effects and any spatial dependence of the potential are neglected in this alternative model.  
Plots of $\delta I_a(V_0)$ and $\delta I_a(\epsilon_a)$ calculated using analytical formula (\ref{eq:iac_p}) are shown by crosses in Figs. \ref{fig:di-dep}(a) and (b). 
These figures reveal that the values of $\delta I_a$ estimated from (\ref{eq:iac_p}) significantly deviate both from the experimental and numerical results -- compare the positions of the circles, solid line and crosses in the figure.  This divergence is not only quantitative, with a scaling factor of at least $\pm 20$, but also qualitative, since the dependence of the theoretical and experimental curves on $V_0$ and $\epsilon_a$ are significantly different. 

This difference highlights that quantum tunneling due to strain waves can only be understood if the propagation of the wave through the material is explicitly included.

\section{Conclusion and Outlook}
\label{sec:con}
In conclusion, we have investigated the DC and time-dependent response of a weakly coupled SL to a hypersonic strain pulse train. In theory and experiment, we have shown that increasing either the applied bias voltage or the amplitude of the acoustic stimuli changes the magnitude of the electric current pulses induced in the SL.  Our quantum model, based on the time-dependent Schr{\"o}dinger equation with a propagating  potential generated by strain stimuli,  reproduces the shape and polarity of current pulses generated in the SL for various intensities of acoustic wave packets and a range of applied bias voltage. Remarkably, the simulations predict current oscillations with a frequency much higher than that of the strain pulse train. 
The quantitative agreement that we report between the theoretical and experimental behavior of the current pulses supports our prediction of these high-frequency oscillations. Our results highlight the potential of SL devices as tunable acoustoelectric transducers for generating high-frequency (sub-THz/THz) electromagnetic waves.

Our quantum model is also applicable to a range of other device architectures, materials, and acoustic wave parameters, and can therefore be applied widely to study 
high-frequency acoustoelectric effects in heterostructures including  novel devices based on 2D and van der Waals materials  \cite{Vdovin2016, Greener2018},
and, even more generally, to cold atom transport in periodic optical lattices driven by a propagating potential \cite{Greenaway2013}.

\begin{acknowledgments}
The work was supported by EPSRC (Grants EP/M016161/1, EP/M016099/1).
We thank Prof. Cunningham, Dr. Hunter, Prof. Davies and Prof. Linfield from the University of Leeds, and Dr. Heywood and Mr. Priestley from Teledyne-e2v Microwave Technology Centre for insightful discussions. \end{acknowledgments}

\section{Appendix: tunnel current in superlattices under photon/phonon excitation}
\label{sec:irr}

We  gain further insight into our results by deriving an analytical model adapted from the theory of nonlinear single-particle tunnel junctions under electromagnetic irradiation \cite{Tucker1979}.   Previously,
we demonstrated that the coherent acoustic waves can be rectified in a SL via a similar process to the rectification of electromagnetic waves in such structures \cite{Poyser2015}. Here, we extend the theoretical framework used in \cite{Poyser2015}  to estimate the magnitude of the electric pulses induced by a strain wave packet in the biased SL.

In a weakly coupled SL, electrons propagate through the lattice via quantum mechanical tunneling between neighboring quantum wells.  Interaction of a SL electron with a coherent strain wave, with frequency $\omega$, will cause the electron to absorb a phonon of energy $\hbar\omega$.  Therefore, assuming tunneling between the quantum wells is independent, i.e. affected only by the properties of adjacent wells, the tunneling rate will be affected only by the energy of the excitation, rather than its phase. This assumption neglects any effects caused by changes of the phase of the acoustic wave as it propagates along the SL. Then the acoustic wave can be described by an effective oscillating electric potential $V_{ac} (t)= (\epsilon_a DL)/(ed\sin \omega t)$ \cite{Tucker1979,Poyser2015}, where $L$ is the SL length. 

If the conductance of the SL varies slowly on the voltage scale $\hbar\omega/e$, and $(\epsilon_a D L)/(\hbar \omega d)$ is small enough, we can estimate the DC and time-dependent current response classically by considering the detection and rectification of an effective voltage signal $V_0+(\epsilon_a DL)/(ed\sin \omega t)$ by a device with a current-voltage characteristic $I_0(V_0)$. In this case,  the amplitude of the time-dependent current component $\delta I_{a}$ can be estimated as \cite{Poyser2015,Tucker1979, Wacker1997b}

\begin{equation}
\label{eq:iac}
\delta I_{a}(V_0,\epsilon_a)=\left(\frac{\epsilon_a D L}{ed}\right) \frac{dI_0(V_0)}{dV_0}.
\end{equation}

Remarkably, the shape of the $I_0(V_0)$ curve is accurately described by a cubic polynomial $I_0(V_0)=aV_0-bV_0^3$, which is widely used to approximate the $N$-shaped current-voltage characteristics of tunnel devices \cite{Chow1964}. A good fit to the data is obtained using the least-squares method [see crosses in Fig. \ref{fig:static}(b)], which gives $a=1.65\times 10^{-2}$ $\text{S} ^{-1}$ and $b=9.1 \times 10^{-3}$ $\text{SV} ^{-2}$. 

Substituting the polynomial approximation of the current-voltage characteristics, $I_0(V_0)=aV_0-bV_0^3$, into Eq. (\ref{eq:iac}) gives

\begin{equation}
\label{eq:iac_p}
\delta I_{a}(V_0,\epsilon_a)=\epsilon_a DL / (ed) (a-3bV^2_0).
\end{equation}
The corresponding dependencies of $\delta I$ on $V_0$ and $\epsilon_a $ are shown by crosses in Fig. \ref{fig:di-dep}.


\begin{thebibliography}{9}

\bibitem{Thomsen1985} C. Thomsen, J. Strait, Z. Vardeny, H J Maris, J. Tauc  and J.J. Hauser,
Coherent phonon generation and detection by picosecond light pulses, Phys. Rev. Lett. \textbf{53} 989 (1984).

\bibitem{Eesley1987}  G.L. Eesley, B.M Clemens and C.A. Paddock, 
Generation and detection of picosecond acoustic pulses in thin metal films, Appl. Phys. Lett. \textbf{50} 717 (1987).

\bibitem{Akimov2017}A.V. Akimov, C.L. Poyser and A. J. Kent, Review of microwave electro-phononics in semiconductor nanostructures, Semicond. Sci. Technol. \textbf{32}, 053003 (2017)

\bibitem{Fowler2008}D. R. Fowler, A. V. Akimov,  A. G. Balanov, M. T. Greenaway, M. Henini, T. M. Fromhold, and A. J. Kent, Semiconductor charge transport driven by a picosecond strain pulse, Appl. Phys. Lett \textbf{92}, 232104 (2008)

\bibitem{Greenaway2009}M.T.Greenaway, A.G.Balanov, D.Fowler, A.J. Kent, T.M. Fromhold, Using sound to generate ultra-high-frequency electron dynamics in superlattices, Microelectronics Journal \textbf{40} 725-727 (2009), doi:10.1016/j.mejo.2008.11.051.

\bibitem{Greenaway2010} M. T. Greenaway, A. G. Balanov, D. Fowler, A. J. Kent, and T. M. Fromhold, Using acoustic waves to induce high-frequency current oscillations in superlattices, Phys. Rev. B \textbf{81}, 235313 (2010)

\bibitem{Apostolakis2017} A. Apostolakis, M. K. Awodele, K. N. Alekseev, F. V. Kusmartsev, and A. G. Balanov, Nonlinear dynamics and band transport in a superlattice driven by a plane wave, Phys. Rev. E \textbf{95} 062203 (2017)

\bibitem{Heywood2016} S.L. Heywood, B.A. Glavin, R.P. Beardsley, A.V. Akimov, M.W. Carr, J. Norman, P.C. Norton, B. Prime, N. Priestley and
A.J. Kent, Heterodyne mixing of millimetre electromagnetic waves and sub-THz sound in a semiconductor device, Sci. Rep. \textbf{6} 30396 (2016)

\bibitem{Kent2006} A. J. Kent, R. N. Kini, N. M. Stanton, M. Henini, B. A. Glavin, V. A. Kochelap, and T. L. Linnik, Acoustic Phonon Emission from a Weakly Coupled Superlattice under Vertical Electron Transport: Observation of Phonon Resonance, Phys. Rev. Lett. \textbf{96}, 215504 (2006)

\bibitem{Batista2007}P. D. Batista, M. Gustafsson, M. M. de Lima Jr.,  M. Beck, V. I. Talyanskii, R. Hey, P. V. Santos, M P. Delsing and J. Rarity, Acousto-electric single-photon detector, Proc. SPIE \textbf{6583}, Photon Counting Applications, Quantum Optics, and Quantum Cryptography, 658304 (2007)

\bibitem{Talyanskii1997}V. I. Talyanskii, J. M. Shilton, M. Pepper, C. G. Smith, C. J. B. Ford, E. H. Linfield, D. A. Ritchie, and G. A. C. Jones, Single-electron transport in a one-dimensional channel by high-frequency surface acoustic waves, Phys. Rev. B \textbf{56}, 15180 (1997)

\bibitem{Cunningham2000} J. Cunningham, V. I. Talyanskii, J. M. Shilton, M. Pepper, A. Kristensen, and P. E. Lindelof, Single-electron acoustic charge transport on shallow-etched channels in a perpendicular magnetic field, 
Phys. Rev. B \textbf{62}, 1564 (2000)

\bibitem{Ahn2007} K. J. Ahn, F. Milde, and A. Knorr, Phonon-Wave-Induced Resonance Fluorescence in Semiconductor Nanostructures: Acoustoluminescence in the Terahertz Range, 
Phys. Rev. Lett. \textbf{98}, 027401 (2007)

\bibitem{Gell2009} J. R. Gell, M. B. Ward, R. J. Young, R. M. Stevenson, P. Atkinson, D. Anderson, G. A. C. Jones, D. A. Ritchie, and A. J. Shields, Modulation of single quantum dot energy levels by a surface-acoustic-wave,
Appl. Phys. Lett. \textbf{93}, 081115 (2008)

\bibitem{Young2011} E. S. K. Young, A. V. Akimov, M. Henini, L. Eaves, and A. J. Kent, Subterahertz Acoustical Pumping of Electronic Charge in a Resonant Tunneling Device, Phys. Rev. Lett. \textbf{108}, 226601 (2011)

\bibitem{Poyser2015a} C.L. Poyser, A. V. Akimov, R. P. Campion and  A. J. Kent, Coherent phonon optics in a chip with an electrically controlled active device,
Scientific Reports \textbf{5}, 8279 (2015)

\bibitem{Brue2012} C. Br\"uggemann, A. V. Akimov, A. V. Scherbakov, M. Bombeck,
C. Schneider, S. H\"ofling, A. Forchel, D. R. Yakovlev, and M. Bayer, Laser mode feeding by shaking quantum dots in a planar microcavity, Nat. Photonics \textbf{6}, 30 (2012)

\bibitem{Wigger2017} D. Wigger, T. Czerniuk, D. E Reiter, M. Bayer and T. Kuhn, Systematic study of the influence of coherent phonon wave packets on the lasing properties of a quantum dot ensemble, New J. Phys. \textbf{19}, 073001 (2017)


\bibitem{Stanton2003} C.J. Stanton, G.D. Sanders, R. Liu, G.W. Chern, C.-K. Sun, J.S, Yahng, Y.D, Jho, J.Y. Sohn, E. Oh, and D.S. Kim, Coherent phonons, nanoseismology and THz radiation in InGaN/GaN heterostructures,
Superlatt. Microstruct. \textbf{34}, 525 (2003)

\bibitem{Armstrong2009} M. R. Armstrong, E. J. Reed, Ki-Y. Kim, J. H. Glownia, W. M. Howard, M. William, E. L. Piner, and J.C Roberts, Observation of terahertz radiation coherently generated by acoustic waves, Nat. Phys. \textbf{5}, 285 (2009)

\bibitem{Poyser2015} C. L. Poyser, A.V. Akimov, A.G. Balanov, R.P. Campion and A. J. Kent, A weakly coupled semiconductor superlattice as a harmonic hypersonic-electrical transducer, New J. Phys. \textbf{17} 083064  (2015)



\bibitem{Akimov2007}A. V. Scherbakov, P. J. S. van Capel, A. V. Akimov, J. I. Dijkhuis, D. R. Yakovlev, T. Berstermann, and M. Bayer, Chirping of an Optical Transition by an Ultrafast Acoustic Soliton Train in a Semiconductor Quantum Well, Phys. Rev. Lett. \textbf{99}, 057402 (2007).



\bibitem{Bloch1929} F. Bloch, About the quantum mechanics of electrons in crystal lattices, 
Journal of Physics {\bf 52},  555-600 (1929)


\bibitem{Esaki1970} L. Esaki and R. Tsu, Superlattice and negative differential conductivity in semiconductors 
IBM J. Res. Dev. \textbf{14}, 61-65 (1970)

\bibitem{Fromhold2004} T. M. Fromhold, A. Patan\`e, S. Bujkiewicz, P. B. Wilkinson,
D. Fowler, D. Sherwood, S. P. Stapleton, A. A. Krokhin, L. Eaves,
M. Henini, N. S. Sankeshwar, and F. W. Sheard, Chaotic electron diffusion through stochastic webs enhances current flow in superlattices,
Nature (London) \textbf{428}, 726-730 (2004)

\bibitem{RENK1998} K.F. Renk and E. Schomburg and A.A. Ignatov and J. Grenzer and S. Winnerl and K. Hofbeck,
Physica B: Condensed Matter \textbf{244}, 00487 (1998), doi:10.1016/S0921-4526(97)00487-0.

\bibitem{Wac02} A. Wacker, Semiconductor superlattices:
a model system for nonlinear transport, Physics Reports {\bf 357},  1–111 (2002)
doi:10.1016/S0370-1573(01)00029-1.

\bibitem{Kastrup1996}J. Kastrup, H.T. Grahn, K. Ploog, R. Merlin, Oscillating electric field domains in  superlattices,
Solid-State Electronics \textbf{40}, 157-160 (1996)

\bibitem{Greenaway2009a} M. T. Greenaway, A. G. Balanov, E. Sch\"oll,  and T. M. Fromhold, Controlling and enhancing terahertz collective electron dynamics in superlattices by chaos-assisted miniband transport
Phys. Rev. B \textbf{80}, 205318 (2009)


\bibitem{Wacker1997b} A. Wacker, S.J. Allen, J.S. Scott, M.C. Wanke and A.-P. Jauho, Possible THz Gain in Superlattices at a Stable Operation Point,
 Phys. Stat. Sol. (b), \textbf{204}, 95-97 (1997)
 
 
\bibitem{Adachi1992} S. Adachi, \textit{Physical Properties of III-V Semiconductor Compounds} (Wiley, 1992).

\bibitem{Gor1993} I. Gorczyca, T. Suski, E. Litwin-Staszewska, L. Dmowski, J. Krupski and B. Etienne, Deformation Potential in High Electron Mobility GaAs/GaAsAs Heterostructures,
Jpn. J. Appl. Phys. \textbf{32} 135 (1993)
 
\bibitem{Balanov2012} A.G. Balanov, M.T. Greenaway, A.A. Koronovskii, O.I. Moskalenko, A.O. Selskii, T.M. Fromhold and A.E. Khramov, The effect of temperature on the nonlinear dynamics of charge in a semiconductor superlattice in the presence of a magnetic field, J. Exp. Theor. Phys. \textbf{114}  836 (2012)
 
 \bibitem{Maksimenko2015} V. A. Maksimenko, V.V. Makarov, A.A. Koronovskii, K.N. Alekseev, A.G. Balanov and A.E. Hramov, The effect of collector doping on the high-frequency generation in strongly coupled semiconductor superlattice, Europhys. Lett. \textbf{109} 47007 (2015)

\bibitem{Alexeeva2012} N. Alexeeva, M. T. Greenaway, A. G. Balanov, O. Makarovsky, A. Patanè, M. B. Gaifullin, F. Kusmartsev, and T. M. Fromhold, Controlling High-Frequency Collective Electron Dynamics via Single-Particle Complexity, Phys. Rev. Lett. \textbf{109}, 024102 (2012)


 
 
\bibitem{Vdovin2016}E.E. Vdovin, A. Mishchenko, M.T. Greenaway, M.J. Zhu, D. Ghazaryan, A. Misra, Y. Cao, S.V. Morozov, O. Makarovsky, T.M. Fromhold, A. Patan\`e, G.J. Slotman, M.I. Katsnelson, A.K. Geim, K.S. Novoselov, and L. Eaves, Phonon-Assisted Resonant Tunneling of Electrons in Graphene Boron Nitride Transistors, Phys. Rev. Lett. \textbf{116}, 186603 (2016)

\bibitem{Greener2018} J. D. G. Greener, A. V. Akimov, V. E. Gusev, Z. R. Kudrynskyi, P. H. Beton,
Z. D. Kovalyuk, T. Taniguchi, K. Watanabe, A. J. Kent, and A. Patan\`{e}, Coherent acoustic phonons in van der Waals nanolayers and heterostructures, Physi. Rev. B \textbf{98}, 075408 (2018)

\bibitem{Greenaway2013}M. T. Greenaway, A. G. Balanov, and T. M. Fromhold, Resonant control of cold-atom transport through two optical lattices with a constant relative speed,
Phys. Rev. A \textbf{87}, 013411 (2013)


\bibitem{Tucker1979}  J. R. Tucker, Quantum limited detection in tunnel junction mixers,
 IEEE J. Quantum Electron. \textbf{15}, 1234-1258 (1979)
 

 \bibitem{Chow1964} W.F Chow, \textit{Principles of Tunnel Diode Circuits} (New York, Wiley, 1964)
 



\end{thebibliography}
\end{document}